\documentclass[conference]{IEEEtran}

\ifCLASSINFOpdf
  \usepackage[pdftex]{graphicx}
  \graphicspath{{../images}}
  \DeclareGraphicsExtensions{.pdf,.jpeg,.png}
\else
  \usepackage[dvips]{graphicx}
  \graphicspath{{../eps/}}
  \DeclareGraphicsExtensions{.eps}
\fi

\newcommand\finline[3][]{\begin{myfont}[#1]{#2}#3\end{myfont}}%
\newenvironment{myfont}[2][]{\csname#2\endcsname[#1]}{}

\usepackage{xspace} 
\newcommand{\framework}{\textsc{\finline[\fontsize{8}{10}]{myFont}{AMIPA}}\xspace}

\newcommand{\method}[1]{\texttt{#1}}
\newcommand{\rd}{\emph{r\&d}\xspace}

\usepackage[usenames, dvipsnames]{color}

\newcommand{\toRemove}[1]{\textcolor{black}{#1}}

\usepackage{listings}
\definecolor{keywordcolor}{rgb}{0.13,0.13,1}
\definecolor{greenComments}{RGB}{63,127,95}
\lstdefinelanguage{ADL}{
    keywordstyle=[2]\color{purple},
    keywordstyle=[3]\color{blue},
    keywords=[2]{boolean, long, Thread, String, double, equals, false, true, to},
    keywords=[3]{System, Component, Connector, Port, Property, Properties, Role, Roles, Ports, Attachments, Attachment, Heuristic, IF, THEN, ELSE, ELSE-IF, RULE, AND},
    morekeywords={to}
    sensitive=false,
    morestring=[b]",
    morecomment=[l]{//},
    }

\lstdefinestyle{adl-style}{
    language=ADL,
    backgroundcolor=\color{white},   
    commentstyle=\color{green},
    numberstyle=\tiny\color{gray},
    basicstyle=\ttfamily\scriptsize,
    breakatwhitespace=false,         
    breaklines=true,                 
    captionpos=b,                    
    keepspaces=true,                 
    showspaces=false,                
    showstringspaces=false,
    showtabs=false,                  
    tabsize=2,
    columns=fullflexible,
    commentstyle=\color{greenComments}, 
}

\lstdefinestyle{rule-style}{
    language=ADL,
    backgroundcolor=\color{white},   
    commentstyle=\color{greenComments},
    keywordstyle=\color{blue},
    numberstyle=\tiny\color{gray},
    basicstyle=\ttfamily\scriptsize,
    breakatwhitespace=false,         
    breaklines=true,                 
    captionpos=b,                    
    keepspaces=true,                 
    numbers=left,                    
    numbersep=5pt,                  
    showspaces=false,                
    showstringspaces=false,
    showtabs=false,                  
    tabsize=2,
    columns=fullflexible,
    rulecolor=\color{black},
    frame=shadowbox,
    numbers=none
}

\lstnewenvironment{adl-listing}[1][]{
    \lstset{style=adl-style, caption=#1}
  }
  {}
  
\lstnewenvironment{rule-listing}[1][]{
    \lstset{style=rule-style, caption=#1}
  }
  {}

\usepackage[table]{xcolor}
\definecolor{Gray}{gray}{0.9}

\usepackage{xspace}
\usepackage{enumitem} 
\usepackage{balance}       
\usepackage{graphics}      
\usepackage[T1]{fontenc}   
\usepackage{txfonts}
\usepackage{mathptmx}
\usepackage[pdflang={en-US},pdftex]{hyperref}
\usepackage{color}
\usepackage{booktabs}
\usepackage{textcomp}
\usepackage{array}
\usepackage{makecell}
\usepackage{listings}

\usepackage{microtype}        
\usepackage{ccicons}          

\begin{document}

\title{Architectural Middleware that Supports Building High-performance, Scalable, Ubiquitous, Intelligent Personal Assistants}

\author{\IEEEauthorblockN{Oscar J. Romero}
\IEEEauthorblockA{Machine Learning Department\\
Carnegie Mellon University}
}

\maketitle

\begin{abstract}

Intelligent Personal Assistants (IPAs) are software agents that can perform tasks on behalf of individuals and assist them on many of their daily activities. 
IPAs capabilities are expanding rapidly due to the recent advances on areas such as natural language processing, machine learning, artificial cognition, and ubiquitous computing, which equip the agents with competences to understand what users say, collect information from everyday ubiquitous devices (e.g., smartphones, wearables, tablets, laptops, cars, household appliances, etc.), learn user preferences, deliver data-driven search results, and make decisions based on user's context.
Apart from the inherent complexity of building such IPAs, developers and researchers have to address many critical architectural challenges (e.g., low-latency, scalability, concurrency, ubiquity, code mobility, interoperability, support to cognitive services and reasoning, to name a few.), thereby diverting them from their main goal: building IPAs.
Thus, our contribution in this paper is twofold: 1) we propose an architecture for a platform-agnostic, high-performance, ubiquitous, and distributed middleware that alleviates the burdensome task of dealing with low-level implementation details when building IPAs by adding multiple abstraction layers that hide the underlying complexity; and 2) we present an implementation of the middleware that concretizes the aforementioned architecture and allows the development of high-level capabilities while scaling the system up to hundreds of thousands of IPAs with no extra effort. 
We demonstrate the powerfulness of our middleware by analyzing software metrics for complexity, effort, performance, cohesion and coupling when developing a conversational IPA.

\end{abstract}


\vspace{-0.3cm}
\section{Introduction}
\vspace{-0.0cm}

%
The advances verified in areas such as natural language processing, semantic web, machine learning, and artificial intelligence, combined with the huge amount of available information made accessible by the Internet, has enabled the creation of Intelligent Personal Assistants (IPAs). 
IPAs are mobile, autonomous, and software agents capable to perform tasks or services on behalf of humans~\cite{Garrido:2010}, from answering general questions to recommend restaurants nearby, hear incoming messages and notifications, get directions, process automated subscription content like weather updates, and customize communications like receipts, shipping notifications, and live automated messages, just to name a few.
%
%
As smartphones computational capabilities increase, IPAs such as Apple Siri, Google Now, Amazon Alexa, Microsoft Cortana, Facebook M, etc. aim to radically disrupt the way people search and consume information on the internet, as well as the manner they communicate and interact with the world.
A relatively new trend in the development of IPAs is to allow the agent to dwell not only in the user's smartphone but also in other devices such as wearables, tablets, personal computers, cars, household appliances, etc., and interact with the surrounding environment. This process can be seen as natural interaction among people, environment, and machines, creating a scenario of ubiquitous computing, that is, a scenario where technology is so integrated with users that they are unaware of the existence of the technological functions that surround them~\cite{Satyan:2001}.
This trend entails several architectural challenges that must be addressed in order to create reliable, efficient, engaging, real-time, and ubiquitous IPAs. Most of these architectural challenges are intrinsically related to architecturally significant requirements such as high-performance, low latency, interoperability, scalability, extensibility, ubiquity, concurrency, mobility, support to reasoning and cognitive services, among others. However, addressing these challenges is a very time consuming, tedious, and burdensome task for researchers and developers (\textbf{\rd} from now on), which divert them from their main goal: building IPA's.
A convenient solution for separating the underlying complexity of the architectural requirements from the IPA agent implementation is the inclusion of a middleware, however, there is a lack of available open-source IPA-oriented middlewares that fulfill all the aforementioned architectural requirements.
Therefore, we propose an IPA \emph{architectural middleware} called \framework (\textbf{A}rchitectural \textbf{M}iddleware for building \textbf{I}ntelligent \textbf{P}ersonal \textbf{A}ssistants), which aims to simplify the development of high-performance, ubiquitous IPAs. 
This paper is organized as follows: Section~\ref{sec_requirements} presents the motivation and related work. Section~\ref{sec_requirements} describes the architecturally significant requirements for building the proposed architecture. Section~\ref{sec_architecture} discusses the architectural design considerations necessary for building the middleware. Section~\ref{sec_implementation} addresses a concrete implementation of the architecture. Section~\ref{sec_evaluation} presents the evaluation of our architectural middleware. Section~\ref{sec_conclusions} summarizes the conclusions and addresses future directions for this work.


\vspace{-0.2cm}
\section{Motivation and Related Work}
\label{sec_motivation}
\vspace{-0.1cm}

\subsection{IPAs Architectures}
\vspace{-0.1cm}
The usage of IPAs was originally promoted by projects like the PAL Program~\cite{PAL:2012} (Personalized Assistant that Learns) proposed by DARPA, with contributions from SRI (the predecessor of SIRI) and several other laboratories with CALO (Cognitive Assistant that Learns and Organizes) project~\cite{Tur:2010}. 
The technological revolution has enabled the creation of new gadgets with higher computing capabilities and new features, such as smartphones. Some smartphones provide a personal digital assistant as one of their main features. Apple's Siri~\cite{Siri:2016}, Google Now~\cite{GoogleNow:2016}, and Microsoft's Cortana~\cite{Cortana:2016} are examples, however, there are little insights about how their system architectures were conceived.
IPAs can normally interact equally with other intelligent objects in the environment (human or machine) to obtain knowledge about different domains. These scenarios are referred to as MAS (multi-agent systems)~\cite{Weiss:1999}. MAS are composed of multiple heterogeneous interactive intelligent agents within an environment, enabling parallel processing inside the system and making it less prone to failures.
Li and Chen~\cite{Li:2009} presented a middleware for an IPA based on MAS and case-based reasoning (CBR) for manufacturing. Their MAS middleware allows the creation of a collaborative environment among intelligent agents, decentralizing the processing in the system.
In~\cite{Fuckner:2014} is proposed an IPA to execute services on behalf of users, using natural language interfaces to process the users' requests, and using OMAS as their multi-agent middleware~\cite{OMAS:2011}. OMAS was designed for building cognitive agents and provides three types of agents: service agents (SA), transfer agents (XA) and personal assistant agents (PA). \toRemove{These agents are organized around a single net local loop and share messages using broadcast mode (UDP). The communication between agents is P2P meaning that there is no central directory, a valuable feature for service-oriented architectures.}
In~\cite{Santos:2016} is proposed an IPA that can be integrated into ubiquitous computing environments in an Internet of Things (IoT) context, by using the CoAP protocol that allows REST architectures to be used in IoT applications, and XMPP message protocol which allows entities to communicate inside a network.
Some other approaches~\cite{Pozna:2013, Paraiso:2005} present some theoretical insights about important aspects to develop IPAs, such as natural language and speech interfaces, but they do not focus on aspects such as the architectural considerations necessary to develop those IPAs.
On the other hand, there some other approaches that use well-defined MAS middlewares such as JADE~\cite{Bellifemine:1999} (Java Agent Development Framework, a FIPA-compliant multi-agent system) to develop and execute IPA agent-based applications.
CoolAgent RS~\cite{Harry:2001} is a context-aware IPA system implemented with JADE that allows contextual information to be freely distributed among agents so that the meaning of that information can be shared and understood. 
Though we initially consider JADE as a middleware for building IPAs, we realized some limitations of this framework, such as the lack of support for latest versions of Android-based devices (e.g, Android 6 and above)~\cite{Weihong:2013, Ughetti:2008} and some issues regarding performance~\cite{Salim:2017}. 
The strongest middleware candidate we took into consideration to develop IPAs was VHT (Virtual Human Toolkit)~\cite{Hartholt:2013}. VHT is a collection of modules, tools, and libraries designed to aid and support researchers with the creation of virtual human conversational assistants. Though its architecture define proper mechanisms for extensibility (through loosely coupled modules) it doesn't scale very well because the latency footprint increases dramatically, there is no proper mechanism to manage the concurrency, and the messaging system is based on a pre-defined configuration of ActiveMQ that doesn't allow to create customized communication patterns. We will discuss how VHT compares to \framework on Section~\ref{sec_evaluation}.

\vspace{-0.2cm}
\subsection{Cognitive Services (CS)}
\vspace{-0.1cm}
CS are a set of REST APIs based on machine learning algorithms intended to create the next generation of applications powered by AI. 
Microsoft Cognitive Services~\cite{MicrosoftCS:2017} enables its Bot Framework and the Cortana (IPA) Intelligent Suite to perform Facial and object Recognition (FR), Automatic Speech Recognition (ASR), Natural Language Understanding and Generation (NLU and NLG), Knowledge Base access (KB), and Semantic Searches (SS). 
IBM Watson~\cite{Watson:2017} provides CS for intelligent searches, deployment of conversational chatbots and virtual agents (NLU/ASR), and visual recognition.
And Google Actions~\cite{Awareness:2017} allows to extend Google Now IPA by providing hooks to conversational CS that trigger actions implemented by developers.
The two most evident disadvantages of these approaches are: 
CS are tight to the IPA platform, e.g., Google actions are only accessed through Google Now IPA and cannot be invoked from other apps; and they cannot be easily replaced by other CS (from other providers), that is, \rd have to re-implement the whole set of CS when switching the provider.

\vspace{-0.3cm}
\section{Requirements}
\label{sec_requirements}
\vspace{-0.1cm}

In this section we describe the \emph{Architecturally Significant Requirements} (ASReqs) that we took into consideration during the development of \framework. According to \cite{Babar:2013}, ASReqs are requirements that: (1) play an important role in determining the architecture; (2) need to be satisfied before the architecture is considered ``stable''; and (3) affect the architecture in measurably identifiable ways. So \framework should address:

\vspace{-0.1cm}
\begin{itemize}[nosep,topsep=1ex,leftmargin=2.0\labelsep]
    \itemsep-0.0em 
    
    \item \emph{ASR01 Interoperability:} it should define common ``contracts'' of how heterogeneous components can exchange data.
    
    \item \emph{ASR02 Multi-session:} it should allow managing multiple simultaneous sessions that share some kind of resources.
    
    \item \emph{ASR03 Scalability:} it should perform efficiently under situations such as an expanding number of connected IPAs.
    
    \item \emph{ASR04 Low-latency:} 
    it should be latency-sensitive, that is, the speed at which the system responds to an asynchronous stream of independent and diverse events that result from interactive user input should not exceed 100 ms \cite{Endo:1996}.
    
    \item \emph{ASR05 Abstraction:} it should exhibit a gradual increase in the level of representation by replacing existing detailed information with information that emphasizes certain aspects important for \rd while hiding the least relevant aspects. As a result, the system's complexity and size should decrease.
    

    \item \emph{ASR06 Distributed and decentralized:} it should define flexible mechanisms to easily build robust distributed IPAs. 
    
    \item \emph{ASR07 Fully portable:} it should allow further implementation of cross-platform multi-language IPAs. 
    

    
    \item \emph{ASR08 Concurrency:} it should exhibit at least the following features proposed by~\cite{armstrong03}: (a) process creation/destruction should be a lightweight operation; (b) asynchronous message passing is the only way for processes to interact; (c) copying message-passing semantics (share-nothing concurrency); (d) processes are strongly isolated; and (e) remote processes appear largely the same as local processes. 
    
    \item \emph{ASR09 Pluggability/Extensibility:} it should provide mechanisms to make IPAs more modular, customizable, and easily extensible. It must allow to extend the functionality of the IPA by easily adding new features and components. 
    
    \item \emph{ASR10 Mobility:} components should be able to run code that is migrated from one machine to another.
    
    \item \emph{ASR11 Event-driven and Service-oriented:} it should allow service management and communication through a loosely-coupled asynchronous messaging mechanism.
        
    
    \item \emph{ASR12 Ubiquity:} it should provide mechanisms to run the middleware in a wide range of devices (e.g., phones, tablets, wearables, etc.) anytime, anywhere and using any format.
    
    \item \emph{ASR13 Reasoninga and Orchestration:} it should reason over the data collected by different devices and orchestrate the interaction between services and components. 
    
    \item \emph{ASR14 Cognitive Services:} it should define reusable contracts for connecting CS that can be replaced any time through a lightweight integration mechanism.
\end{itemize}
\vspace{-0.2cm}
We have identified the most critical requirements to create our architectural middleware, however, there are so many other requirements that we are not covering for now (e.g, security, privacy, maintainability, etc.) but expect to do it in the future.

 \vspace{-0.2cm}
\section{Architectural Model} \label{sec_architecture}
 \vspace{-0.2cm}
 
\begin{figure}
\includegraphics[width=\columnwidth] 
{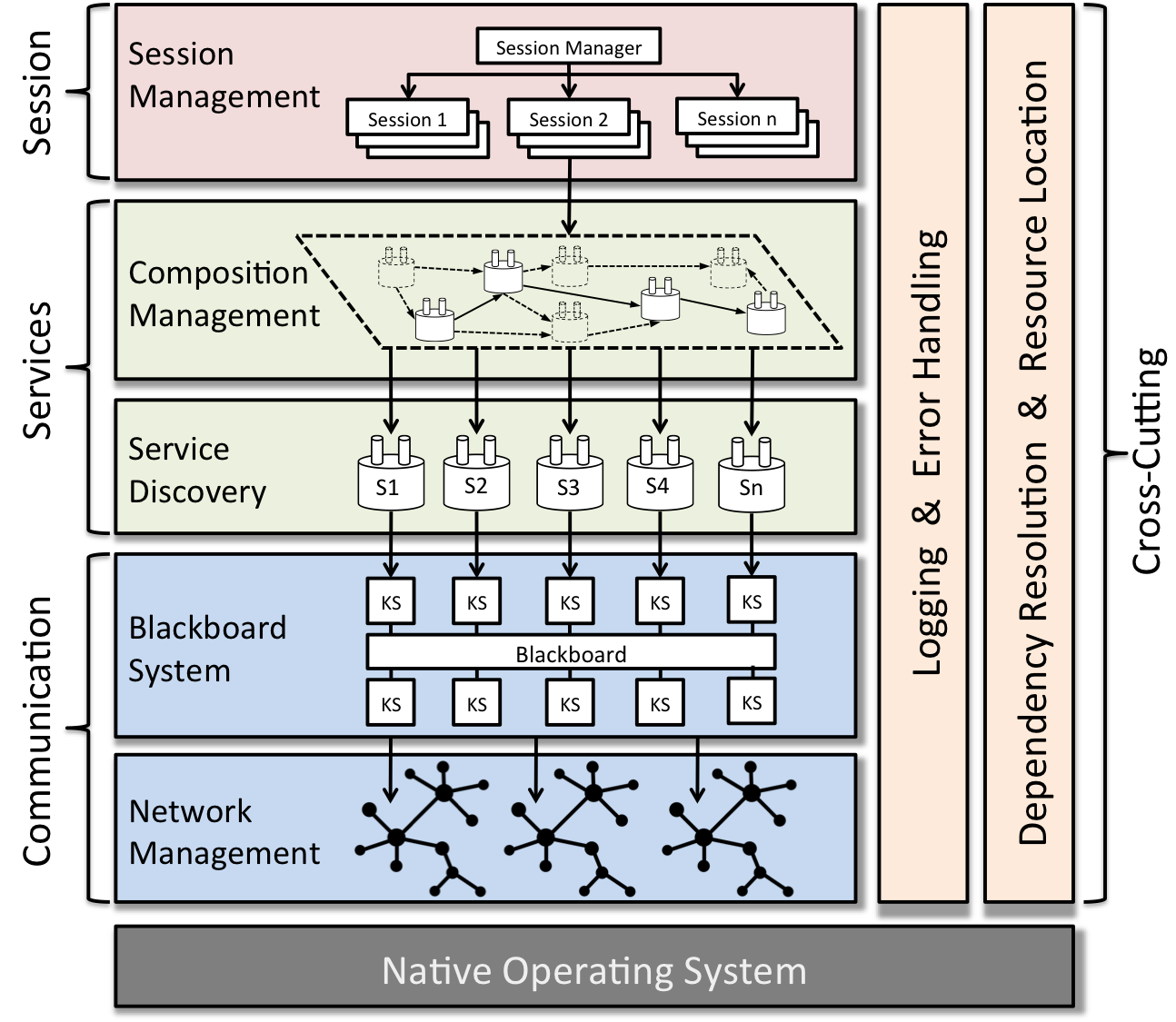}
 \vspace{-0.7cm}
\caption{Overall Architecture. KS: Specialist Knowledge Source}
\label{fig_architecture}
\vspace{-0.7cm}
\end{figure}
%
%
\framework's architecture follows the basic principles of a Clean Architecture~\cite{Martin:2017} guaranteeing better separation of concerns and better modularization of the IPA project. \toRemove{In a nutshell, the Clean Architecture defines concentric circles that represent different areas of software, where the outer circles are mechanisms and the inner circles are policies. The overriding rule that makes this architecture work is the Dependency Rule which says that source code dependencies can only point inwards, nothing in an inner circle can know anything at all about something in an outer circle.} Following this rule, \framework's layers can only point from the upper layers towards the bottom layers and not the contrary. 
Figure~\ref{fig_architecture} depicts a high-level architecture for \framework with four identifiable macro layers: Communication, Decision Making, Management, and Cross-Cutting. 
Though the model resembles a monolithic architecture in the sense that the design is self-contained and their components seem to be interdependent rather than loosely coupled, it is not, this is only a high-level representation of the system's architecture, which is appropriate for our purpose of capturing the vocabulary. On the contrary, \framework is inspired by the microservices architectural style. According to~\cite{Dragoni:2017,Fowler:2014,Richardson:2017}: microservices is an architectural style that breaks an application down into smaller, cohesive, independent, loosely coupled, elastic, resilient, and scalable collection of services, each running in its own process and communicating with lightweight mechanisms. \framework is not purely oriented to microservices, only some layers support this style, as we describe later.
In software development, \emph{architecture} and \emph{middleware} are different concepts, while a software architecture define the system's high-level behaviors such as components, connectors, communication rules, connection rules, etc.; middleware provides an implementation, generally based on services, that sits between an application and the OS or the programming language, and where the architectural concerns are secondary. According to~\cite{Nenad:2003}, the relationship between architecture and middleware and their respective shortcomings suggest the possibility of coupling architecture modeling and analysis approaches with middleware technologies in order to get the best of both worlds. 
Next, we describe the structural and behavioral viewpoints of \framework and some architectural views. The concerns to be addressed were described in section~\ref{sec_requirements} and stakeholders for all viewpoints is the \rd team. 
\subsubsection{Structural Viewpoint}
\framework components communicate with each other through architecture-level software connectors that are implemented later using a middleware. This approach preserves the properties of the architecture-level connectors while leveraging the beneficial capabilities of the underlying middleware. 
Listing 1 presents an excerpt of \framework's architectural description using ACME-ADL~\cite{Garlan:2000}. This oversimplified description will be enough to identify the main architectural elements of \framework. Let's focus on the most relevant components of this architecture description. There are six components, one represents the client (i.e., a smartphone, smartwatch, tablet, laptop, etc.), two are in charge of managing the user sessions (Session and SessionManager), and the other three are responsible for decision making (ProcessOrchestrator, PluggableComponent, and CognitiveService), as we will discuss in detail in further sections. 
There are also three main connectors in charge of linking different components as described in the Attachments block: the Broker connector allows the IPA-client component to connect to both SessionManager and Session components; the Blackboard connector mediates the communication between a Session component and a ProcessOrchestrator component, and between the ProcessOrchestrator and a set of PluggableComponents; and the CommunicationController connector, which supports multiple communication protocols, allows the communication between a PluggableComponent and a CognitiveService. 
Finally, there are defined two architectural constraints in terms of Heuristics (a Heuristic constraint is taken to be a rule that should be observed, but may be selectively violated), one for low-latency and another for scalability. The first heuristic states that the total latency (the sum of all the latencies for the decision making components) should ideally not exceed 100 ms, as proposed in section~\ref{sec_requirements}. The second heuristic says that when the system scales up to one million sessions, it should still keep the latencies in the range of 100-120 ms (i.e, the performance should not be affected when scaling the system up). 
%
%
%
%
%
%
\begin{adl-listing}[AMIPA's architecture description using ACME-ADL]
System AMIPA = {
    Component IPA-client = { // smartphone, smartwatch, tablet, etc.
        Ports {connect; send-request, disconnect} }
    Component SessionManager = { // it controls sessions' lifecycle
        Ports {open-session; close-session};
        Properties {sessions : Session[];
                    multithreaded : boolean = true} }
    Component Session = {
        Ports {process; send-response};
        Properties {orchestrator : ProcessOrchestrator;
                    multisession : boolean = true} }
    Component ProcessOrchestrator = {  
        Ports {post, send-response};
        Properties {components : PluggableComponents[];
                    latency : long;
                    thread : Thread = new Thread} }
    Component PluggableComponent = {  
        Ports {execute; post};
        Properties {service : CognitiveService;
                    latency : long;
                    thread : Thread = new Thread} }
    Component CognitiveService = {
        Ports {execute; return-result};
        Properties {latency : long = 80 << units="ms">>;
                    run-locally : boolean : false;
                    type-of-service <type = "KB | ASR | NLU |..."> ;
                    endpoint : String : "";
                    thread : Thread = new Thread} }
    Connector Broker = {
        Roles {sender; receiver};
        Properties {asynchronous : boolean = true;
            type-of-request = ["ASR-request", "NLU-request", ... ];
            service-directory =  // registry for service discovery
                sender.ASR -> receiver.lookup(name, ASR)
                sender.NLU -> receiver.lookup(name, NLU)
                sender.NLG -> receiver.lookup(name, NLG) ...;
            protocol : socket-TCP} }
    Connector Blackboard = {  
        Roles {publisher; subscriber};
        Properties {asynchronous : boolean = true;
                    subscribers : BlackboardListener[]} }
    Connector CommunicationController = {
        Roles {caller; callee};
        Properties {asynchronous : boolean = true;
                    protocol : <type = "INPROC | IPC | TCP | RPC">}}
    ...
    Attachments {
        IPA-client.connect to Broker.sender;
        SessionManager.open-session to Broker.receiver;
        IPA-client.send-request to Broker.sender;
        Session.receive-request to Broker.receiver;
        Session.process to Blackboard.publisher;
        ProcessOrchestrator.on-execute to Blackboard.subscriber;
        ProcessOrchestrator.process to Blackboard.publisher;
        PluggableComponent.on-execute to Blackboard.subscriber;
        PluggableComponent.post to CommunicationController.caller;
        CognitiveService.execute to CommunicationController.callee }
    Property total-latency : long =  (ProcessOrchestrator.latency + PluggableComponent.latency + CognitiveService.latency);
    Heuristic total-latency <= 100; //constraint (ms)
    Heuristic (size(SessionManager.sessions) <= 1,000,000) and (total-latency >= 100) and (total-latency <= 120);
}
\end{adl-listing}
%

\subsubsection{Communication Viewpoint (ASR01, ASR05, ASR06, ASR07, ASR10, ASR11, ASR12)}

\begin{figure}
\centering
\includegraphics[width=2.5in] 
{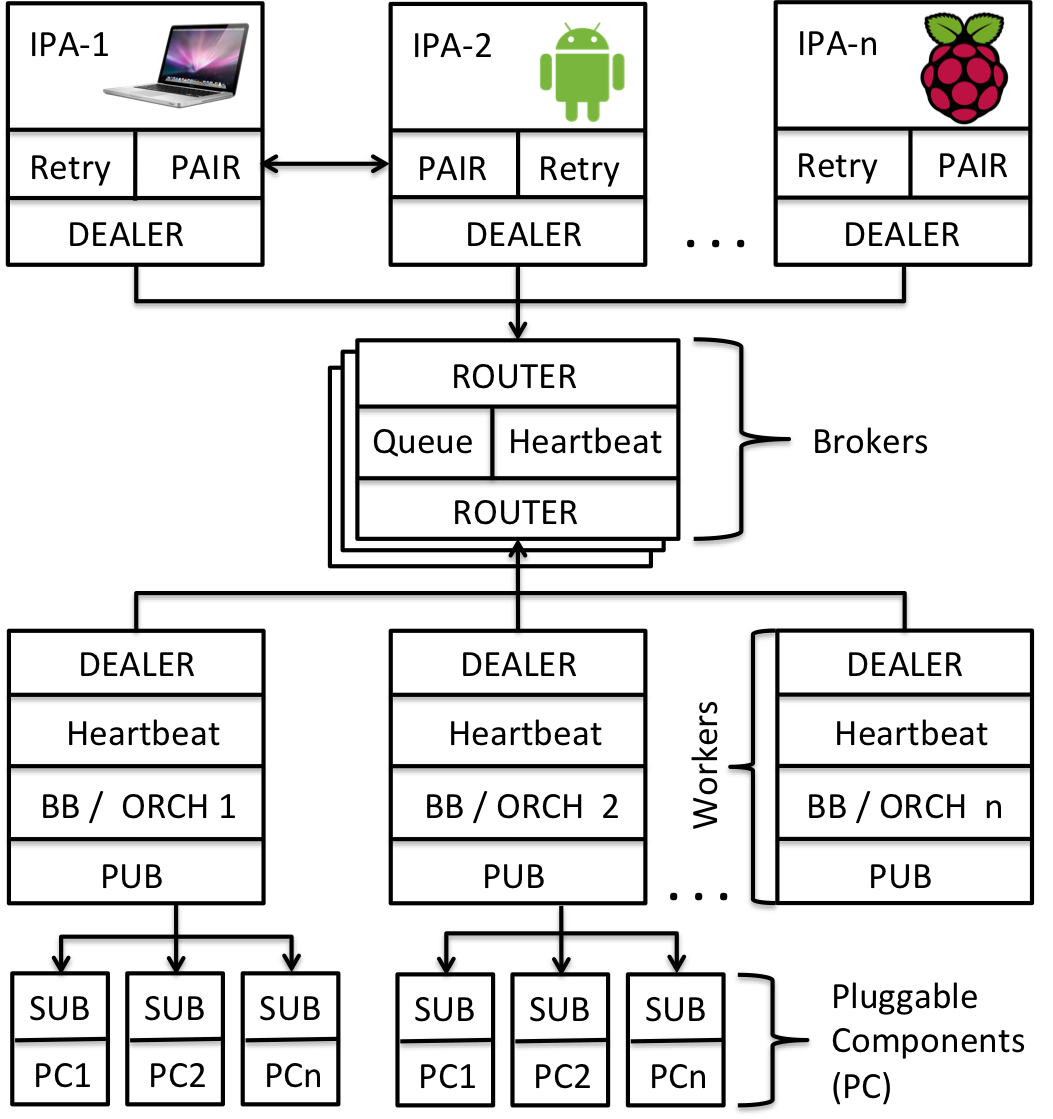}
\vspace{-0.2cm}
\caption{Communication model. BB: Blackboard}
\vspace{-0.7cm}
\label{fig_topology}
\end{figure}

\emph{Messaging:} We propose an extension of the Asynchronous Majordomo Specification \cite{Hintjens:2017}, a \emph{Service-Oriented Reliable Queuing} pattern that defines a dialog between a set of client applications (those issuing requests, e.g., an IPA client on user's phone), one or multiple broker devices (for routing, service discovery, task distribution, etc.), and a pool of worker applications (those processing the requests which can register for specific services (e.g., an ASR service), as depicted in figure~\ref{fig_topology}. 
%
%
The DEALER and ROUTER patterns allow \framework to send multiple requests and multiple replies simultaneously and asynchronously, which means that workers are not passive modules that only listen for IPA client requests, but active and autonomous processes that can proactively communicate with IPA clients anytime. 
%
%
\emph{Blackboards} use a PUB-SUB pattern to broadcast messages to specific \emph{pluggable components}. IPA apps can communicate with each other either directly through a PAIR-PAIR pattern or indirectly through a broker. 
It is worth mentioning that all communication patterns in this architecture (i.e., ROUTER, DEALER, PAIR, PUB/SUB) are design-level constructs that realize architecture-level connectors, meaning that they define common contracts that can be further realized using different \framework implementations. 
This layer supports communication between co-located services trough INPROC protocol (thread to thread within a single process),  IPC protocol (messages between local processes), TCP protocol (data exchange between IPA applications), and RPC (remote procedure calls that allow cognitive services to collaborate).
The messaging layer is based on the following microservices patterns~\cite{Richardson:2017}: \emph{Messaging}, \emph{Domain-specific Protocol}, and \emph{Remote Procedure Invocation} patterns for inter-service communication, and \emph{Circuit Breaker} pattern for reliably handling partial failure.
%
%
%
%
%

\emph{Blackboard System:} A blackboard system is better explained through the following metaphor~\cite{Corkill:1991}: a group of specialists work cooperatively to solve a problem by watching the blackboard and looking for an opportunity to apply their expertise to the developing solution. Such opportunities arise when an event occurs (a change is made to the blackboard) that enables the specialist to act. 
%
%
This process continues until the problem has been solved. \framework's blackboard system is composed of three main modules: (1) Pluggable Components (or Specialist Knowledge Sources according to the metaphor), independent modules that contain the knowledge needed to solve the problem and can be widely diverse in representation and inference techniques; (2) a global working memory (the blackboard) containing input data, partial solutions, and other data that are in various problem-solving states; and (3) a Process Orchestrator (the control module in the metaphor) that keeps problem solving on track to insure that all crucial aspects of the problem are receiving attention, and to balance the stated importance of different specialist's contributions.
Pluggable Components (PC) subscribe to specific messages with the blackboard, so one PC may subscribe to multiple messages and multiple PCs may subscribe to the same message. On the other hand, the Process Orchestrator (PO) intercepts all the messages broadcasted by the blackboard. The blackboard receives messages from two sources: from external components (e.g., IPA apps, remote services, etc.) by using the underlying communication layer, and from internal components through an event-driven mechanism (i.e., PCs that post a partial solution of the current problem). When a new message is received, the blackboard notifies all the subscribers about that specific message, and then subscribers may extract the data that come wrapped in a BlackboardEvent object. 
%
%
Although multiple PCs may be activated by the same message, it is the PO who decides which PC can post to the blackboard.
%
%
The Blackboard define a set of connectors that allow heterogeneous PCs to interact with each others through message-passing mechanisms. This layer uses the \emph{Messaging} pattern of the microservices style.
%
\subsubsection{Concurrency Viewpoint (ASR03, ASR04, ASR08)}
Processes that runs on single threads do faster when compared to multi-threaded processing, because it involves no context switching and no synchronization/locking. Therefore, in order to build a really efficient concurrency model, \framework does not use mutexes, locks nor semaphores meant to orchestrate the parallel processing. Instead, each object lives in its own thread, and threads do not share objects and only communicate through an inter-process protocol.

%
\subsubsection{Pluggability and Extensibility Viewpoint (ASR03, ASR05, ASR09, ASR14)} 
\label{sec_pluggable}
\begin{figure}
\includegraphics[width=\columnwidth] 
{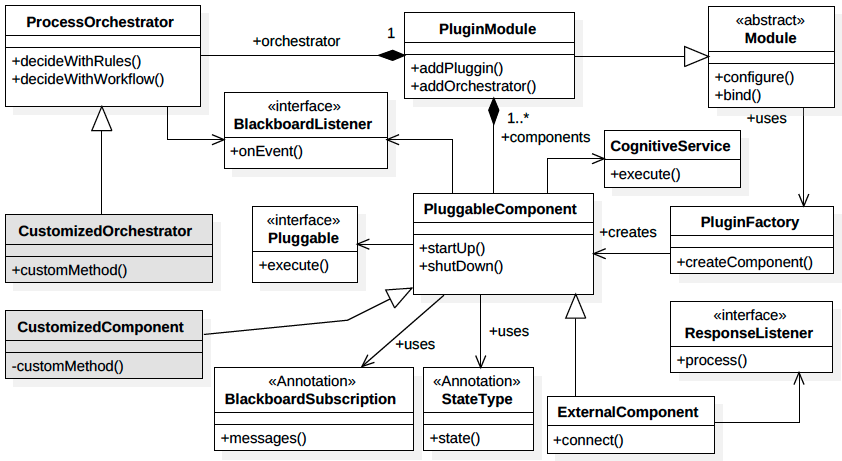}
\vspace{-0.6cm}
\caption{Simplified Class Diagram for the Decision-Making layer}
\vspace{-0.1cm}
\label{fig_plugins}
\end{figure}

Pluggability is an essential feature in \framework, which allows \rd team to build IPAs that are modular, customizable, and easily extensible. 
In \framework, Pluggable Components (PC) are organized around IPA capabilities, that is, they encapsulate the interaction with self-contained cognitive services such as ASR, NLU, NLG, KB, etc. that may run either locally or remotely. When running locally and within the same process boundary, PCs behave as components (i.e, units of software that are independently replaceable and upgradeable~\cite{Fowler:2014}) that communicate through inexpensive in-process calls, whereas when crossing process boundaries, they behave as (micro)services that use well-defined remote interfaces to exchange information among distributed components using a mechanism such as a web service request or remote procedure call. Implementing a PC as a component or a service is left at \rd team discretion and depends on IPA's context. 
In Figure~\ref{fig_plugins} is depicted a view that illustrates the semantics of the Pluggability and Extensibility Viewpoint. The abstract class \emph{PluggableComponent} exposes some common functionality, so \rd can then link against \framework in their applications and include the appropriate implementation by extending this abstract class. It is worth mentioning that the \emph{PluggableComponent} class is the first-class construct of \framework, it defines a set of functions and mechanisms for registering callback functions for different types of messages, defines lifecycle methods for starting-up and shutting-down the component/service, and implements both the \emph{Pluggable} and \emph{BlackboardListener} interfaces which define the contracts of how PCs should behave when using either a direct-invocation or an event-driven approach. 
In the former approach, \rd are responsible for calling each component in the desired order (synchronously or asynchronously) by invoking the \method{execute()} method through a \emph{ProcessOrchestrator}. 
On the other hand, for the latter approach, the PC's \method{onEvent()} method is automatically triggered each time the blackboard is modified (e.g., insertion and deletion of elements). 
%
Moreover, in order to simplify the implementation of a PC, subclasses of \emph{PluggableComponent} may use annotations for both subscribing to specific messages (\emph{BlackboardSubscription} annotation) and defining the PC state (\emph{StateType} annotation). More specifically, the behavior of a PC is determined by any of these three types of states: (1) \emph{Stateful} PCs keep a representation of user's state, so a new PC is instantiated each time a session is created for each user; (2) \emph{Stateless} PCs do not keep the user's state, so they are instantiated as singletons; and (3) \emph{Pool} PCs that can create a fixed number of instances that split the work among them through a load-balancing algorithm. 
Additionally, the \emph{Module} class defines high-level abstraction methods for the runtime creation, configuration and binding of a set of PCs. Now, the \emph{PluginModule} class exposes methods to register different kind of PCs and orchestrators in an internal registry (for service discovery). 
Equally important is the \emph{PluginFactory} class, which dynamically instantiates specific PCs contained in the PluginModule's registry by using reflection and dependency injection. 
%
%
If a PC behaves as a local component then it only has to extend the \emph{PluggableComponent} class, as \emph{CustomizedComponent} class does, otherwise, if it behaves as a remote service, it registers with the PluginModule class and an \emph{ExternalComponent} instance is automatically created to serve as a proxy between \framework and the remote service. The ExternalComponent class implements the ResponseListener interface, so when asynchronous responses from remote services are received by \framework, then the callback \method{process()} is invoked. 
The idea behind this viewpoint is that \rd can build high-level IPA functionality on top of PCs without having to worry about where their PCs will be running on, that is, it is completely transparent for them whether PCs are running locally or as remote services.
\setlength{\textfloatsep}{5pt}
%
\subsubsection{Orchestration and Reasoning Viewpoint (ASR05, ASR13, ASR14)} \label{sec_orchestrator}

Process Orchestrators are in charge of arbitrating the interactions and dynamics among PCs. As we discussed on Section \ref{sec_pluggable}, \framework's orchestrators provide two types of control of execution: direct-invocation and event-driven. When using an \emph{event-driven} approach, the execution of components is mostly based on events that are triggered by the blackboard system; then events are intercepted by the orchestrator, which will decide if any kind of data transformation is needed; and finally, these events are delivered by the orchestrator to the corresponding PC. Conversely, the \emph{direct-invocation} allows \rd to explicitly define the way that components may interact with each other. For that purpose, \framework provides two mechanisms to define the control of execution: a \emph{workflow controller} (\method{decideWithWorkflow()} method) and a \emph{rule-based} system (\method{decideWithRules()} method). For the former mechanism, \rd can design interaction workflows using Petri net-like semantics, that is, directed bipartite graphs where nodes are PCs and edges are triggering conditions that define the transitions among those PCs. For the latter mechanism, \rd can create production rules expressed as IF-THEN structures, where conditions correspond to a set of premises in the blackboard that must be true, and actions are execution triggers such as posting new  premises on the blackboard, invoking one or several components, etc. 
Listing 2 presents a rule-based orchestration script that controls the interaction among 7 PCs (and therefore Cognitive Services) for a user-IPA conversational interaction. In a nutshell, there are at least two main scenarios that the orchestration process can cover: 1) an ideal scenario where user speaks, the voice is recorded and processed by ASR, then NLU finds a corresponding user-intent for user's utterance, next the Dialogue Manager (DM) generates a system-intent (a system action), then NLG generates a system utterance, and finally TTS (Text-To-Speech) converts the text into a spoken voice output (rule sequence: [01, 02, 03, 04, 05]); and 2) an exceptional scenario where NLU cannot match the user's utterance with a user-intent (i.e., confidence value is low), so the system asks the user for clarification, next if the system cannot find a proper user-intent yet then it passes the user utterance to the Question\&Answering component (QA), which in turn if does not find an answer then passes the user utterance to a chatbot powered by crowdsourcing (CCS), and finally, TTS sinthesizes the CCS' output (rule sequence: [01, 02, 03, 05, 01, 02, 03, 06, 07]). 
The reasoning process is not only performed by the orchestrator but also by Cognitive Services that can use diverse kind of reasoning mechanisms depending on its own nature (e.g., a Semantic Search CS typically uses semantic inference to extract new facts from a KB).

%

\begin{rule-listing}[Rules for a conversational interaction]
RULE-01: Activate ASR
  IF     Blackboard.event equals MIC_Event 
  THEN   PC.ASR.execute(MIC_EVENT.bytes) 
RULE-02: Activate NLU
  IF     Blackboard.event equals ASR_Event 
  THEN   PC.NLU.execute(ASR_EVENT.utterance) 
RULE-03: Activate DM
  IF     Blackboard.event equals NLU_Event THEN
    IF      NLU_Event.user_intent.confidence > 0.7 
  	THEN    PC.DM.execute(NLU_EVENT.user_intent)
  	ELSE-IF   NLU_Event.already_asked_confirmation equals false
  	  THEN    PC.NLG.post(NLG_EVENT.ASK_FOR_CLARIFICATION)
  	  ELSE    PC.QA.execute(ASR_Event.utterance)
RULE-04: Activate NLG
  IF     Blackboard.event equals DM_Event 
  THEN   PC.NLG.execute(DM_EVENT.system_intent)
RULE-05: Activate NLG-TTS
  IF     Blackboard.event equals NLG_Event 
  THEN   PC.TTS.execute(NLG_EVENT.utterance)
RULE-06: Activate QA-TTS
  IF     Blackboard.event equals QA_Event THEN
    IF      QA_Event.response not-equals null
    THEN    PC.TTS.execute(QA_EVENT.answer)
    ELSE    PC.CCS.execute(ASR_Event.utterance)
RULE-07: Activate CCS-TTS
  IF     Blackboard.event equals CCS_Event
  THEN   PC.TTS.execute(CCS_Event.answer) 
\end{rule-listing}
\vspace{-0.2cm}

%


\subsubsection{Dependency Resolution and Service Discovery Viewpoint (ASR05, ASR09, ASR11)} \label{sec_dependency}

In order to support pluggability and flexibility features, \framework uses two common design patterns that assemble components from different sources into a cohesive application: a \emph{Resource Locator} and a \emph{Dependency Injector}. Both components decouple system implementation from its dependencies, so these dependencies can be replaced or updated with no change to the system implementation \cite{Fowler:2002}. 
%
The \emph{Resource Locator} is a singleton registry that contains references to different kind of \framework's resources (e.g., components, remote services, databases, etc.) and encapsulates the logic that locates them.
On the other hand, the \emph{Dependency Injector} removes internal dependencies from \framework by allowing dependent objects to be injected into the classes/methods by an external caller. The \emph{Dependency Injector} provides automatic instantiation and lifecycle management of classes the developer registers such as \emph{pluggable components}, \emph{orchestrators}, \emph{sessions}, etc.
Both the Resource Locator and the Dependency Injector are mechanisms for flexibility that maximizes dependency on interfaces while minimizing dependency on specific implementation. \toRemove{Consequently, \framework can support ``pluggable'' implementation classes that are used depending on the circumstances.}
The components in this viewpoint are inspired by the \emph{microservice chasis}, \emph{externalized config}, \emph{log aggregation}, \emph{service registry}, and \emph{distributed tracing} microservice patterns~\cite{Richardson:2017}.
\subsubsection{Session Management Viewpoint (ASR02, ASR03, ASR12)} \label{sec_session_mgmt}
\begin{figure}[t]
\includegraphics[width=\columnwidth] 
{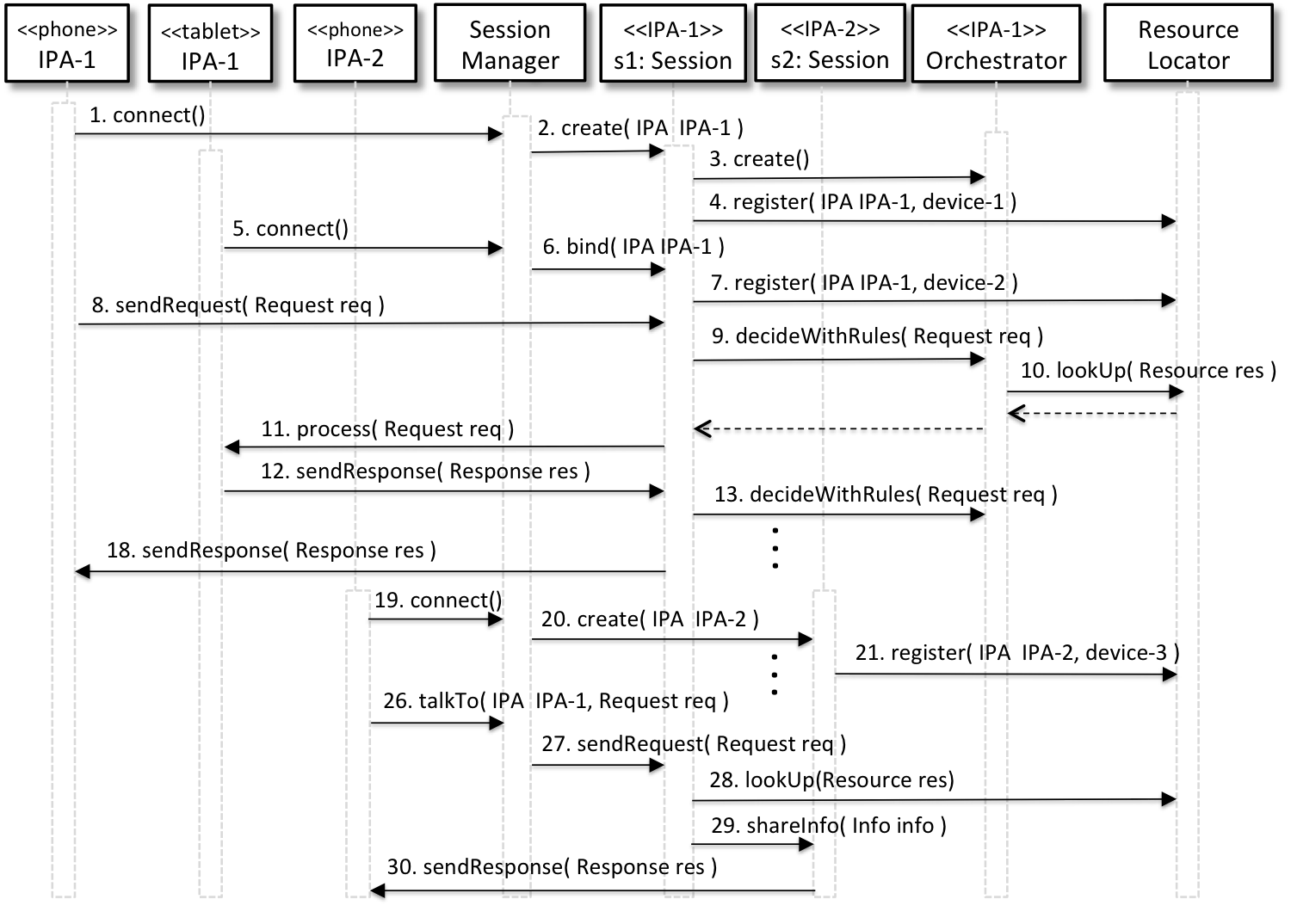}
\vspace{-0.6cm}
\caption{Sequence Diagram for Session Management Layer}
\vspace{-0.1cm}
\label{fig_session}
\end{figure}


%
This viewpoint focuses on the \emph{``multi-user''}, \emph{``session-sharing''} and \emph{``cross-IPA task coordination''} capabilities of \framework. 
The \emph{multi-user} capability is possible thanks to the concurrency and extensibility properties of \framework's underlying layers, which allow the system to be easily scaled up to hundred thousands IPAs instances with no extra development effort and minimal latency footprint. 
%
The \emph{session-sharing} and the \emph{cross-IPA task coordination} capabilities are accomplished by two fundamental components: \emph{Session} and \emph{SessionManager}. 
There exist a unique session per user/IPA that carries out the following tasks: (1) it keeps the state of user's interaction; (2) it controls session inactivity timeouts and decides whether to automatically reconnect itself; (3) it processes all messages sent from different user's devices and then passes them to the corresponding process orchestrator; and (4) it provides a mechanism to guarantee that multiple ubiquitous user devices communicate through the same session so they can share information and collaborate in a distributed, inexpensive and robust fashion, supporting this way the session-sharing capability of the system.
On the other hand, the SessionManger is responsible for: (1) handling the entire life-cycle of sessions (i.e., create, destroy, pause, resume, reconnect, etc.); (2) moderating the interaction among multiple IPAs so they can share information and collaborate to perform tasks, supporting this way the cross-IPA task coordination capability; 
and (3) controlling the interaction of multiple instances of \framework, that is, developers may define a master/slave configuration where a master \framework instance manage all the IPAs sessions, whereas one or multiple slave \framework instances may be deployed on a distributed environment to perform specific computations and therefore avoid overloading the master instance. 
Figure \ref{fig_session} shows an example of how multiple IPAs, co-existing in multiple ubiquitous devices, may interact with each others. Assume there are two users, Alice and Bob, who live in the same apartment. Bob's IPA (\emph{IPA-1}) co-exists in two devices, a phone and a tablet, and a new session is created for \emph{IPA-1} when Bob connects to his phone. When subsequent \emph{IPA-1} instances are connected (e.g., Bob's tablet IPA) they are bound to the existing session 
(steps 1-7 in Figure~\ref{fig_session}). 
Then, Bob wants to check the weather app on his phone, which requires his location to filter the forecasting results, however, the location option is turned-off on his phone but turned-on on his tablet. Thus, the weather PC running on the \emph{IPA-1} phone instance sends a request through the SessionManager to the location PC running on the \emph{IPA-1} tablet instance, which in turn finds out Bob's current location and returns this value to the weather PC on the phone. All this interaction among IPA instances is completely transparent for Bob (steps 8-18). 
After some time, Alice opens her IPA on her phone, and a new session is created (\emph{IPA-2}) when her IPA connects to \framework. Alice and Bob need to coordinate who will do the groceries and what needs to be shopped. \emph{IPA-1} shares Bob's location, who is closer to a grocery store, however, it is Alice who has the shopping list, so \emph{IPA-2} shares the list with \emph{IPA-1} and both IPAs (i.e., their process orchestrators) come to the conclusion that Bob should do the shopping. All this coordination and resource sharing is done through the interaction between Sessions, services and the SessionManager (steps 19-30).

\vspace{-0.2cm}
\section{Implementation} \label{sec_implementation}
\begin{figure}
\includegraphics[width=\columnwidth] 
{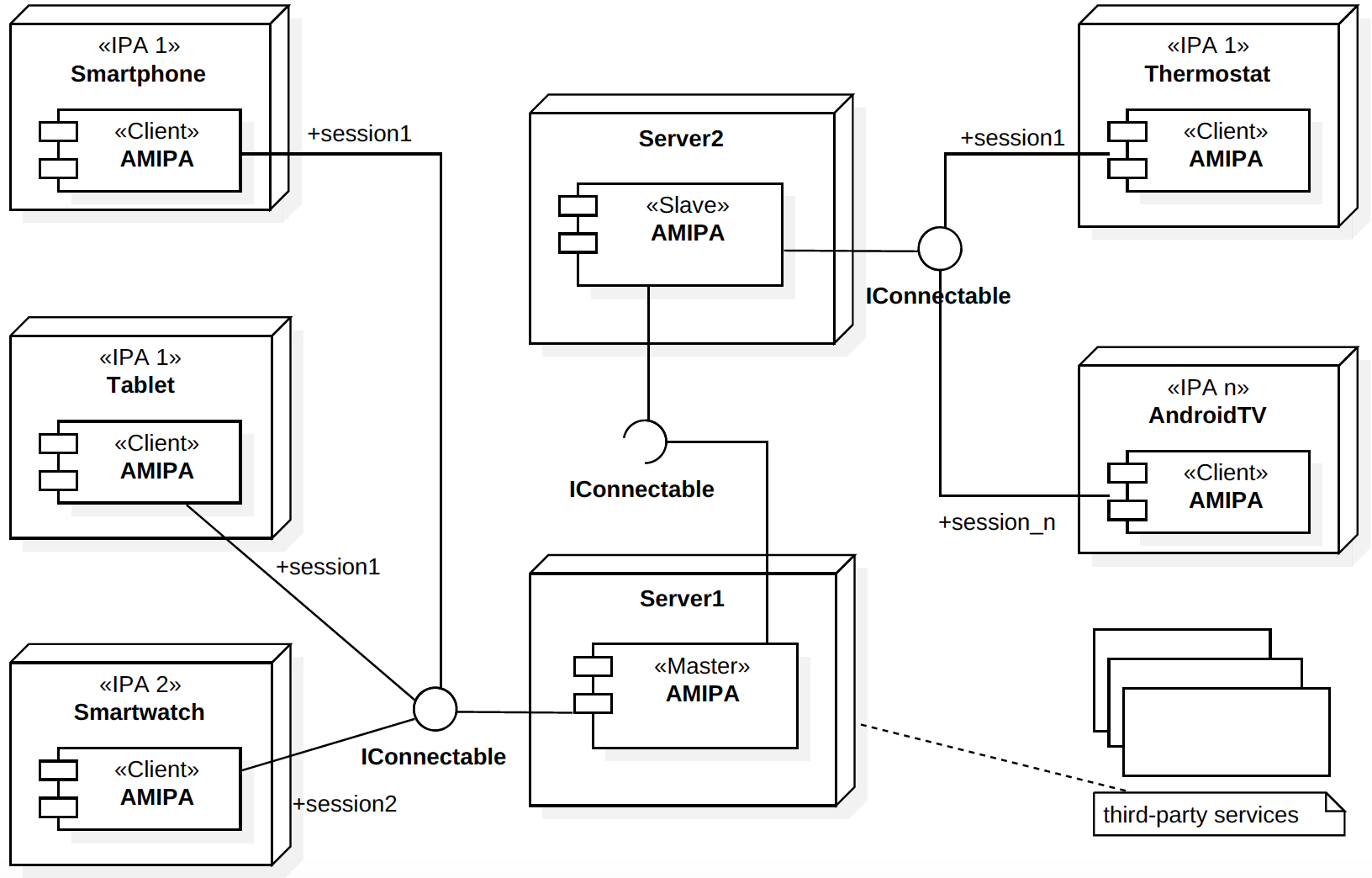}
\vspace{-0.5cm}
\caption{Illustration of an AMIPA's Physical View.}
\label{fig_deploy}
\end{figure}

\vspace{-0.1cm}
In this section, we provide some details of \framework's \emph{middleware} implementation\footnote{See our GitHub repo: \textcolor{blue}{\url{https://github.com/ojrlopez27/multiuser-framework}}}, that is, the prefabricated structure that deals primarily with scaffolding, meaning that it provides pre-built modules that are easy to use/extend and, therefore, releasing \rd from having to create them from scratch. 
We decided to implement the concurrency and messaging layers by using ZeroMQ (ZMQ)~\cite{Veber:2011}, a high-performance asynchronous messaging library aimed at use in distributed and concurrent applications with minimal latency footprint. There are some bench-markings \cite{Veber:2011, Dworak:2011} that demonstrate why ZMQ performs significantly better than other messaging/concurrency frameworks. 
The selection of ZMQ also aligns with the microservices principle of building \emph{smart endpoints and dump pipes}~\cite{Fowler:2014}, that is, messaging should be implemented over a lightweight message bus, where the infrastructure chosen should be typically dumb (dumb as in acts as a message router only, and implementations such as ZeroMQ fits well because they don't do much more than provide a reliable asynchronous fabric) while the smarts still live in the endpoints that are producing and consuming messages: in the services.
%
Furthermore, ZMQ enabled us to address several non-functional requirements such as: 
(a) it allows a wide range of ubiquitous devices (e.g., phones,  tablets, raspberry  pi's,  Microcontrollers,  etc.) to connect to \framework thanks to it defines a socket-based API that supports different kind of unicast and multicast protocols (INPROC, IPC, TCP, TIPC, PGM, EPGM, NORM, SOCKS5) as well as M2M communication; 
(b) its non-blocking nature dramatically minimizes message delivery latency by dispatching, delivering and queuing messages in parallel to the regular processing performed by the system; 
(c) thanks to ZMQ, \framework is highly inter-operable, portable, cross-language and cross-platform, so it is possible to connect either a 32KB embedded chip or a z/OS mainframe running IBM dialects of Unix, using any of the most known programming languages \cite{Sustrik:2016}. According to~\cite{Meng:2017, Karagiannis:2014}, even IoT applications can get access to ubiquitous data in rich sensing pervasive environments by using a M2M messaging mechanism based on ZMQ; 
(d) it uses CurveZMQ, a protocol for secure messaging across the Internet; and
(e) since ZMQ uses the same API for inter-machine, inter-process and inter-thread communication, \framework can scale up seamlessly from a one-process-per-core scenario to a grid fabric of thousands of machines. 
Based on the foregoing, the proposed ZMQ-based communication topology (see Figure~\ref{fig_topology}) enables \framework to scale to large numbers (thousands) of workers (services) and clients (IPA apps), where a single broker thread can switch several million messages per second, and multi-threaded implementations (offering multiple virtual brokers, each on its own port) can scale to as many cores as required. 
%
We implemented \framework's middleware on both Java and Python programming languages and defined a set of custom annotations (in Java) and decorators (in Python), that serve as syntactic metadata that ease and simplify the implementation of system modules and components. 
Pluggable Components can communicate with each others using two mechanisms: an even-driven approach based on ZMQ and a mobile-code approach based on Mobility-RPC~\cite{Gallagher:2016}, a Java library that allows pluggable components to autonomously migrate from one host to a different host during their execution, and spontaneously update IPAs (running on ubiquitous devices) with new functionality or context-dependent program code.
For message formatting we used protobuf~\cite{Protobuf:2017}, a Google's language-neutral, platform-neutral, extensible mechanism for serializing structured data.
The Dependency Resolution layer was implemented using Google Guice~\cite{Berlin:2015}, an open source generic framework for dependency injection using Java annotations, which also supports AOP (Aspect Oriented Programming). Using guice, \rd can easily and quickly extend \framework's features by alleviating the need for factories and promoting the creation of injectors and method interceptors that efficiently and unobtrusively perform cross-cutting logic (e.g., logging, error handling, etc.). 
We also used Google Guava~\cite{Joshua:2016}, a set of core libraries and API's to control the sessions lifecyle and build the orchestrator's workflows, among other things. The former functionality is carried out by a ServiceManager component that monitors session's state transitions, and the latter uses guava's graphs, a library for modeling graph-structured data where nodes correspond to pluggable components and the edges between them correspond to conditional transitions.
For the orchestrator's rule-based control of execution mechanism, we used the Java Rule Engine API, a light and fast rule engine in compliance with the JSR94 specification.  
Figure~\ref{fig_deploy} illustrates an example of how \framework can be deployed on a distributed environment. \framework provides a server library tha can be deployed on one or multiple host computers (e.g, Server 1 and Server 2) and a lightweight client library that can be deployed in a wide range of ubiquitous devices (e.g, smartphones, tablets, smartwatches, TVs, thermostats, etc.). If the user has installed multiple IPA instances on different devices (let's say IPA-1 which is installed in a smartphone, a tablet, and a thermostat) then they will share the same session (session-1) to send and receive messages to/from different components and services. While the server library support all layers, the client library only supports the concurrency, messaging, pluggable components, cross-cutting and (partially) session management layers.

\vspace{-0.2cm}
\section{Evaluation} \label{sec_evaluation}
\vspace{-0.2cm}

In this section we present an empirical metric-based comparison between \framework and VHT (as we mentioned in Section~\ref{sec_motivation}, VHT was our initial candidate as an IPA middleware before deciding to build our own middleware).
To this purpose, we compared the implementation of an IPA using both middlewares. In a nutshell, the IPA was a conversational agent that generated different kind of recommendations (e.g., movies, news, etc.) while keeping user's engagement through social dialogue. The IPA had access to 7 different CS deployed on 3 remote servers: Google ASR, Multisense for non-verbal behavior recognition~\cite{multisense:2017}, Microsoft NLU, NLG, Dialogue Manager~\cite{socially:2016}, Social Intention Recognition~\cite{zhao:automatic}, and a Social Reasoner for making decisions about conversational strategies~\cite{romero:2017}).
%
\subsubsection{Measuring Latency and Performance} \label{sec_performanceanalysis}
In order to measure the latency of both \framework and VHT, we conducted an experiment\footnote{Servers hardware configuration: EC2 t2.2xlarge AWS instance, Canonical, Ubuntu, 16.04 LTS, 8 cores, amd64 xenial image build on 2017-04-14} where we measured the round trip time (in milliseconds) for 1,000 messages to flow through all 7 CS and for the reply to be received, using $10^1$, $10^2$, $10^3$, $10^4$, and $10^4$ IPA clients. Each experiment was performed 10 times, and then their harmonic mean (which mitigates the impact of large outliers) was calculated. Finally, we calculated the speed in milliseconds per message sent (ms/msg) and the performance improvement rate between both approaches, which was estimated using the equation: $(VHT - \framework)/VHT$. The results are shown in Table~\ref{latency_comparison}.
In general, \framework's performance surpassed VHT's performance in a high rate (from 58.62\% to 95.15\%). It is worth noting that, when scaling from 10 to $10^5$ IPAS, \framework's latency had very little footprint in comparison with VHT, going from $12\times10^{4}$ milliseconds per message (with 10 IPAs) to $27\times10^{4}$ ms/msg (with $10^5$ IPAs), while VHT went from $29\times10^{4}$ to $566\times10^{4}$ ms/msg.
%
%
%
We also ran a one-way ANOVA test to analyze the difference among the means of the 2 groups of data, and given that the obtained p-value was less than the significance level ($p = 0.0049 \leq 0.5$) then we could conclude that there was a statistically significant difference between \framework and VHT results.
The performance experiments revealed a clear correlation between the number of IPAs and the performance rate of \framework vs. VHT: the more messages were sent the better \framework performed in comparison with VHT. This is due to the optimization of \framework's concurrency and communication modules: we drastically reduced the message-passing latency by using low-level thread manipulation and zero-copy techniques instead of using the typical concurrency model based on locks, mutexes, and semaphores.
\begin{table}[t]
  \caption{Latency Comparison. AMIPA vs. VHT}
  \vspace{-0.3cm}
  \label{latency_comparison}
  \centering
  \begin{tabular}{ | r |  r |  r |  r |  r |  r | }
  \hline
    IPAs & AMIPA & AMIPA & VHT & VHT & Performance\\
            & (ms) &  (ms/msg) & (ms) & (ms/msg) & (Rate \%)\\ \hline
    $10^1$     &   12   &  0.0012 & 29 & 0.0029 & 58.62\%  \\ \hline
    $10^2$     &   147  & 0.0015 & 738 &  0.0074 & 80.08\%  \\ \hline
    $10^3$     &   1,895 & 0.0019 & 19,867 & 0.0199  & 90.46\%   \\ \hline
    $10^4$     &   24,604 & 0.0025 &  328,975 & 0.0329  & 92.52\%  \\ \hline
    $10^5$     &   27,4681 & 0.0027 &  5,665,592 & 0.0566  & 95.15\%  \\ \hline
  \end{tabular}
\vspace{-0.5cm}
\end{table}
%
\subsubsection{Measuring Abstraction, Pluggability and Extensibility} \label{sec_metrics}
\begin{table}[t]
\caption{Metrics Comparison of AMIPA vs. VHT.}
\vspace{-0.3cm}
  \label{abstraction_comparison}
  \centering
  \begin{tabular}{ | l | r | r | r | }
    \hline
    Metric & AMIPA & VHT & Performance Rate \\ \hline
    CC	            & 5.2 	&	12.6	&	58.73\% \\ \hline
    MHF (\%)		& 63.87	& 	35.62	&	79.30\% \\ \hline
    AHF (\%)		& 95.32	&	89.12	&	6.95\% \\ \hline
    ILF + EIF (FP)	& 17 	&   17    	&   0\%	\\ \hline
    EI + EO + EQ (FP)	& 15	&   23  	&   34.78\%			\\ \hline
    TFP (FP)		& 182	&   345 	&   47.24\%			\\ \hline
    EPD (person/day)   & 391.30 	&   741.75	    &	47.24\%		\\ \hline
    CBO		        & 3.44		&	3.78	&	8.99\% \\ \hline
    CF (\%)	        & 12.76 	    &   28.43	&	55.11\% \\\hline
    LCOM	        & 0.63		&	2.84	&	77.81\% \\ \hline
  \end{tabular}
\end{table}
There exist different approaches to measure software abstraction, but the general consensus is that the more abstract an application is, the less complex and effort-consuming its development is~\cite{Glass:2002,Damasevicius:2006}. So our initial hypothesis was: \framework should significantly decreased the complexity, size and effort to build the proposed scenario in comparison with VHT. 
For complexity we used the \emph{Cyclomatic Complexity} metric (CC), defined as the number of linearly independent paths within a graph that represents the source code flows, and is calculated as: $M = E - N + 2P$, where $E$ is the number of edges of the graph, $N$ is the number of nodes, and $P$ is the number of connected components~\cite{McCabe:1976}. We used a software metrics tool called MetricsReloaded~\cite{Metrics:2017} and collected the results shown in Table~\ref{abstraction_comparison}. We deduced that both \framework and the VHT have low complexity (according to~\cite{McCabe:1976}, high complexity is over 15), however, the improvement rate demonstrated that \framework reduces the complexity on $\cong59\%$ in comparison with VHT. 
We also measured the level of abstraction in terms of the minimum amount of implementation details that were exposed to the developer without loosing information content. 
To this purpose, we used two MOOD metrics (Metrics for Object Oriented Design)~\cite{Abreu:1996}: the \emph{Method Hiding Factor} (MHF) and the \emph{Attribute Hiding Factor} (AHF) which were calculated across all classes in the system. It is worth noting that while \framework improved method hiding in a $\cong79\%$, it only improved attribute hiding in a $\cong7\%$, which is not particularly a significant difference between the two middlewares. 
In terms of size metric, we used Function Points (FP), a widely accepted industry standard (ISO/IEC 20926:2009) for functional sizing. 
FP are estimated in terms of the amount and functional complexity of data functions ( i.e., Internal Logical Files ILF and External Interface Files EIF); and transaction functions (i.e., External Inputs EI, External Outputs EO, and External Inquiries EQ). 
Based on the results, we observed that the main difference between both middlewares was an increment of 8 FP's (for transaction functions) in VHT, which means that \framework reduced the amount of transactional functionality to be developed in $\cong35\%$. 
The Total Function Point measure (TFP) 
reflects that \framework is $\cong47\%$ smaller in functionality (this means that less functionality has to be implemented to meet the same system's requirements) 
in comparison to VHT, which in turn represents a drop in effort in the same proportions. 
The Effort Person/Day (EPD)  estimation is computed as $EPD = TFP/DR*DPM$, where DR is the delivery rate (in average, a java developer can implement 10 FP's per month~\cite{isbsg:2017}) and DPM is days per person-month (21.5 business days per month). 
%
EPD can be better understood in terms of time and number of persons required to develop the IPA, let's say we have a team of 5 persons, using VHT it would take 148.4 days (741.75/5) while using \framework would take 78.2 days (391.3/5), which means a reduction of $\cong47\%$ of the required effort when using our approach. However, our empirical study revealed that development effort can be reduced more than 60\% when using \framework, and this discrepancy with EPD may be due to unconsidered elements during the estimation.
%
Measuring pluggability and extensibility can be achieved by calculating the amount of coupling and cohesion in the system: the more loosely-coupled and high-cohesive the system is the more pluggable and extensible. 
We used 3 different metrics~\cite{Chidamber:1994}: Coupling Between Object classes (CBO), Lack of Cohesion Methods (LCOM), and Coupling Factor (CF). In CBO, two classes are coupled when methods declared in one class use methods or instance variables of the other class; LCOM defines the number of different methods within a class that reference a given instance variable; and CF is the ratio of the maximum possible number of couplings in the system to the actual number of couplings not imputable to inheritance. The results are presented in Table~\ref{abstraction_comparison}. 
According to~\cite{Chidamber:1994} a CBO $>$ 14 is too high, therefore both approaches have loose coupling, however, using CBO we demonstrated that \framework only reduced coupling in $\cong6\%$. We obtained a more significant difference between both approaches when using CF, this time, \framework reduced the coupling at a rate of $\cong55\%$. 
Our approach lessens coupling due to the use of dependency injection patterns and event-driven communication instead of direct invocations to classes. 
In LCOM metric, a result equals to 0 indicates a cohesive class, higher than 0 indicates that the class needs or can be split into two or more classes. 
Therefore, the results suggest that both approaches have certain lack of cohesion, however, \framework seems to be $\cong78\%$ more cohesive than VHT. The main reason of this improvement is due to the architectural class decomposition into high-cohesive classes with a clear separation of concerns (e.g., communication, service discovery, components, orchestrators, etc.). Nevertheless, the presented metrics are, in some way, biased and extremely sensible to the style of programming.

\section{Conclusions and Future Work}
\label{sec_conclusions}

In this paper, we have presented an architectural middleware solution called \framework that supports the construction of high-performance, distributed, ubiquitous, and scalable Intelligent Personal Assistants. The solution consisted of two parts: 1) a universal architecture that exposes the most critical design aspects, by using different levels of abstraction and design patterns, that can be further concretized using any programming language; and 2) a concrete high-performance, high-scalable middleware implementation using a Java-based approach. 
\framework hides the underlying complexity of the environment; insulates the IPA applications from explicit protocol handling, disjoint memories, data replication, network faults, and parallelism; masks the heterogeneity of operating systems, programming languages, networking technologies, and distributed ubiquitous devices to facilitate IPA programming and management; and provides a better abstraction that allows researchers and developers to build IPAs with less code and less programming errors.
Our future work will focus on several aspects: we are planning to make \framework completely opensource so the developer and research community can take advantage of its powerfulness. Next steps will be to use software standards for modularity, extensibility and pluggability such as OSGi (Open Service Gateway initiative) or a fully-oriented microservice architecture style. Also, we have identified the need of creating a semantic layer on top of middleware to improve the service discovery process, provide more accurate and relevant information to higher-level layers, and make inferences about user's context.

\section*{Acknowledgment} 
This research was supported in part by Yahoo! and Verizon through the CMU-Yahoo InMind project.

%
%
%
%
%

\bibliographystyle{IEEEtran}
\bibliography{main}
\vspace{-0.2cm}

\end{document}